\documentclass[12pt]{article}

\begin{document}
\input amssym.tex

\title{Covariant representations of the de Sitter isometry group}

\author{Ion I. Cot\u aescu \thanks{E-mail:~~cota@physics.uvt.ro}\\
{\it West University of Timi\c soara,}\\{\it V. Parvan Ave. 4,
RO-300223 Timi\c soara}}

\maketitle

\begin{abstract}
We show that the induced representations of the de Sitter isometry group proposed many years ago by Nachtmann are equivalent to those derived from our general theory of external symmetry. These methods complete each other leading to a coherent theory of covariant fields with spin on the de Sitter spacetime. Some  technical details of these representations are presented here for the first time.   

Pacs: 04.20.Cv, 04.62.+v, 11.30.-j
\end{abstract}

\vspace*{12mm} Keywords: de Sitter isometries; covariant fields;
induced representations; discrete isometries.

\newpage
\section{Introduction}

The theory of quantum fields with spin on curved local-Minkowskian manifolds can be correctly constructed only in orthogonal (non-holonomic) local frames where the half-integer spins do make sense \cite{WALD,SG}. Then the entire theory must be tetrad-gauge covariant in order to not affect its physical meaning. Since the isometries can change this gauge we proposed to enlarge the concept of isometry considering external symmetry transformations that preserve not only the metric but the tetrad-gauge too \cite{ES}. 

We defined thus the {\em covariant} representations (CR) of the isometry groups which transform the components of the fields with spin in local frames. We have shown that the CRs are  {\em induced}  by the (non-unitary) finite-dimensional representations of the $SL(2,\Bbb C)$ group which is the universal covering group of the gauge group (of the Minkowski metric), $L^{\uparrow}_{+}\subset SO(1,3)$ \cite{ES}. Obviously, the scalar CR reduces to the well-known natural representation.  The generators of these CRs are the differential operators produced by the Killing vectors (associated to isometries) according to the  generalized Carter and McLenagan formula \cite{CML,ES,EPL}. 

Our approach is helpful on the de Sitter  spacetime where all the free field equations can be analytically solved while the specific  $SO(1,4)$ isometries offer us  conserved observables with a well-defined physical meaning \cite{CCC}. In this framework we derived the quantum modes of the usual free fields with spin in various frames of the de Sitter spacetime  performing the canonical quantization in different representations \cite{C}. 

In the FLRW chart of conformal time of the de Sitter manifold, the Dirac quantum modes correctly normalized in momentum representation were found many years ago by Nachtmann \cite{Nach} which considered another method of constructing covariant representations (NCR).  Moreover, he obtained the generators in momentum representation of the spinless unitary representations of the isometry group indicating how the covariant scalar and Dirac fields can be quantized in canonical manner. 

As a matter of fact, we follow the same line attempting to construct a coherent quantum field theory on the de Sitter spacetime even though sometimes our results differ from those found by other authors. This is because the CRs (or NCRs) are less used in literature as long as one prefers to work with linear representations \cite{Gaz} which are better studied \cite{UIR} or to consider directly two-point functions avoinding thus the quantization procedures. These functions were derived either applying the harmonic analysis on pseudo-orthogonal Lie groups \cite{Dobrev} or, simply, by looking for maximally symmetric two-point solutions of the field equations \cite{Wood}. On the other hand, many cosmological models ignore the spin using only scalar fields transforming according to the natural representation. 

Under such circumstances,  we believe that it is interesting to investigate the relation between the NCRs and our CRs.  The main purpose of this paper is to demonstrate that both these methods of constructing induced CRs are equivalent in the de Sitter case, up to some conventions and notations. In addition, we show that the Nachtmann approach represents a useful tool that completes the framework of our CRs. We have thus the opportunity to discuss some specific properties of these representations analysing the structure of their transformations produced by the de Sitter isometries.

The paper is organized as follows. In the second section we briefly review the geometric context of the local-Minkowskian spacetimes following to present in the third one the construction of the induced CRs resulted from our  theory of external symmetry \cite{ES}. In section 4 we discuss how the de Sitter isometries are related to the linear transformations of the five-dimensional flat spacetime in which the de Sitter hyperboloid is embedded. The next section is devoted to the Nachtmann method of constructing CRs \cite{Nach} pointing out that this is equivalent to our general approach applied  to the Sitter case \cite{ES,CCC}. In section 6 we give some examples showing for the first time how the continuous and discrete de Sitter isometries give rise to the concrete transformations of the CRs. Our concluding remarks are presented in the last section.   

\section{Preliminaries}

Let $(M,g)$ be a local-Minkowskian spacetime equipped with {\em local}  frames $\{x;e\}$ formed by a local chart (or natural frame) $\{x\}$ and a non-holonomic orthogonal frame $\{e\}$. In a given local chart of coordinates $x^{\mu}$, labelled by natural indices, $\mu, \nu,...=0,1,2,3$, the orthogonal frames and the corresponding coframes, $\{\hat e\}$, are defined by the tetrad fields $e_{\hat\mu}$  and  $\hat e^{\hat\mu}$, which are labelled by local indices, $\hat\mu, \hat\nu,...=0,1,2,3$, and obey the usual duality, $\hat e^{\hat\mu}_{\alpha}\,
e_{\hat\nu}^{\alpha}=\delta^{\hat\mu}_{\hat\nu}$, $ \hat
e^{\hat\mu}_{\alpha}\, e_{\hat\mu}^{\beta}=\delta^{\beta}_{\alpha}$,
and orthonormalization,  $e_{\hat\mu}\cdot e_{\hat\nu}=\eta_{\hat\mu
\hat\nu}$, $\hat e^{\hat\mu}\cdot \hat e^{\hat\nu}=\eta^{\hat\mu
\hat\nu}$, conditions. These fields define the local derivatives $\hat\partial_{\hat\alpha}=e^{\mu}_{\hat\alpha}\partial_{\mu}$
and the basis 1-forms $\tilde\omega^{\hat\alpha}(x)=\hat e^{\hat\alpha}_{\mu}(x)dx^{\mu}$. The metric tensor $g_{\mu
\nu}=\eta_{\hat\alpha\hat\beta}\hat e^{\hat\alpha}_{\mu}\hat
e^{\hat\beta}_{\nu}$ raises or lowers the natural indices while for
the local ones  we have to use the flat metric
$\eta=$diag$(1,-1,-1,-1)$ of the Minkowski spacetime $(M_0,\eta)$
which is the pseudo-Euclidean model of $(M,g)$. 

The metric $\eta$ remains invariant under the transformations of the
group $O(1,3)$  which includes the Lorentz group,
$L_{+}^{\uparrow}$, whose universal covering group is $SL(2,\Bbb
C)$.  In the usual covariant parametrization, with the real parameters, $\omega^{\hat\alpha \hat\beta}=-\omega^{\hat\beta\hat\alpha}$, the transformations 
\begin{equation}
A(\omega)=\exp\left(-\frac{i}{2}\omega^{\hat\alpha\hat\beta}
S_{\hat\alpha\hat\beta}\right) \in SL(2,\Bbb C) 
\end{equation}
depend on the covariant
basis-generators of the $sl(2,\Bbb C)$ Lie algebra, $S_{\hat\alpha\hat\beta}$, which are the principal spin operators generating all the spin terms of other operators. In this parametrization  the matrix elements in local frames of the transformations $\Lambda(\omega)\equiv \Lambda[A(\omega)]\in L_{+}^{\uparrow}$ associated to $A(\omega)$ through the canonical homomorphism  can be expanded as
$\Lambda^{\hat\mu\,\cdot}_{\cdot\,\hat\nu}(\omega)= \delta^{\hat\mu}_{\hat\nu}
+\omega^{\hat\mu\,\cdot}_{\cdot\,\hat\nu}+\cdots$. Obviously, $\Lambda(0)=I$ is 
the identity transformation of $L_{+}^{\uparrow}$.

Assuming that $(M,g)$ is orientable and time-orientable we can
restrict ourselves to consider $G(\eta)=L^{\uparrow}_{+}$  as the
gauge group of the Minkowski metric $\eta$ \cite{WALD}. This is the
structure group of the principal fiber bundle whose basis is $M$ while the group ${\rm Spin}(\eta)=SL(2,\Bbb C)$  represents the structure group of the spin fiber bundle \cite{WALD,SG}. In this framework one can build the gauge-covariant (or, simply, covariant) field theories whose physical meaning does not depend on the local frames one uses. 

The {\em covariant fields}, $\psi_{(\rho)}:\, M\to {\cal V}_{(\rho)}$, are locally defined over $M$ with values in the vector spaces ${\cal V}_{(\rho)}$ carrying the finite-dimensional non-unitary representations $\rho$ of the group $SL(2,\Bbb C)$. In general, these representations are reducible being equivalent to direct sums of irreducible ones, $(j_+,j_-)$ \cite{WKT}. They determine the form of the covariant derivatives of the field $\psi_{(\rho)}$ in local frames,
\begin{equation}\label{der}
D_{\hat\alpha}^{(\rho)}= e_{\hat\alpha}^{\mu}D_{\mu}^{(\rho)}=
\hat\partial_{\hat\alpha}+\frac{i}{2}\, \rho(S^{\hat\beta\, \cdot} _{\cdot\,
\hat\gamma})\,\hat\Gamma^{\hat\gamma}_{\hat\alpha \hat\beta}\,.
\end{equation}
The local connection coefficients   
$\hat\Gamma^{\hat\sigma}_{\hat\mu \hat\nu}=e_{\hat\mu}^{\alpha}
e_{\hat\nu}^{\beta}(\hat e_{\gamma}^{\hat\sigma}
\Gamma^{\gamma}_{\alpha \beta} -\hat e^{\hat\sigma}_{\beta,
\alpha})$  assure the covariance of the whole theory under the (point-dependent) tetrad-gauge transformations, $\tilde\omega \to \Lambda[A]\tilde\omega$ and $\psi_{(\rho)}\to  \rho(A)\psi_{(\rho)}$, produced by the automorphisms $A\in SL(2,{\Bbb C})$ of the spin fiber bundle.

\section{Covariant representatios}

Let us assume now that $(M,g)$ has isometries, $x\to x'=\phi_g(x)$, given by the (non-linear) representation $g\to \phi_g$ of the isometry group $I(M)$ defined by the composition rule $\phi_g\circ \phi_{g'}=\phi_{gg'}$, $\forall g,g'\,\in I(M)$. Then we denote by $id=\phi_e$ the identity function, corresponding to the unit $e\in I(M)$, and deduce $\phi_g^{-1}=\phi_{g^{-1}}$. In a given parametrization, $g=g(\xi)$ (with $e=g(0)$), the isometries
$x\to x'=\phi_{g(\xi)}(x)=x+\xi^a k_a(x) +...$ lay out the Killing vectors $k_a=\partial_{\xi_a}\phi_{g(\xi)}|_{\xi=0}$ associated to the parameters $\xi^a$ ($a,b,...=1,2...N$).  

In general, the isometries may change the relative position of the local frames affecting thus the physical interpretation. For this reason we proposed the theory of external symmetry \cite{ES} where we introduced the combined transformations $(A_{g},\phi_{g})$  able to correct the position of the local frames. These transformations must preserve not only the metric  but the tetrad-gauge too, transforming  the 1-forms as $\tilde\omega(x')=\Lambda[A_{g}(x)]\tilde\omega(x)$. Hereby, we deduce  \cite{ES},
\begin{equation}\label{Axx}
\Lambda^{\hat\alpha\,\cdot}_{\cdot\,\hat\beta}[A_{g}(x)]= \hat
e_{\mu}^{\hat\alpha}[\phi_{g}(x)]\frac{\partial
\phi^{\mu}_{g}(x)} {\partial x^{\nu}}\,e^{\nu}_{\hat\beta}(x)\,,
\end{equation}
assuming, in addition, that  $A_{g=e}(x)=1\in SL(2,\Bbb C)$. We obtain thus the desired transformation laws, 
\begin{equation}\label{es}
(A_{g},\phi_{g}):\qquad
\begin{array}{rlrcl}
e(x)&\to&e'(x')&=&e[\phi_{g}(x)]\,,\\
\psi_{(\rho)}(x)&\to&\psi_{(\rho)}'(x')&=&\rho[A_{g}(x)]\psi_{(\rho)}(x)\,.
\end{array}
\qquad
\end{equation}
that preserve the tetrad-gauge. We have shown that the pairs $(A_{g},\phi_{g})$ constitute a well-defined Lie group that can be seen as a representation of the universal covering group of $I(M)$ denoted here by $S(M)$ \cite{ES}. 

In a given parametrization, $g(\xi)$, for small values of  
$\xi^{a}$, the $SL(2,\Bbb C)$ parameters of $A_{g(\xi)}(x)\equiv
A[\omega_{\xi}(x)]$ can be expanded as
$\omega^{\hat\alpha\hat\beta}_{\xi}(x)=
\xi^{a}\Omega^{\hat\alpha\hat\beta}_{a}(x)+\cdots$, in terms of the
functions
\begin{equation}\label{Om}
\Omega^{\hat\alpha\hat\beta}_{a}\equiv {\frac{\partial
\omega^{\hat\alpha\hat\beta}_{\xi}} {\partial\xi^a}}_{|\xi=0}
=\left( \hat e^{\hat\alpha}_{\mu}\,k_{a,\nu}^{\mu} +\hat
e^{\hat\alpha}_{\nu,\mu}
k_{a}^{\mu}\right)e^{\nu}_{\hat\lambda}\eta^{\hat\lambda\hat\beta}
\end{equation}
which  are skew-symmetric,
$\Omega^{\hat\alpha\hat\beta}_{a}=-\Omega^{\hat\beta\hat\alpha}_{a}$,
only when $k_a$ are Killing vectors \cite{ES}.

The last of equations (\ref{es}) defines the CRs  {induced} by the finite-dimensional ones, $\rho$, of the group $SL(2,\Bbb C)$. These are operator-valued representations, $T^{(\rho)} \,:\, (A_{g},\phi_{g})\to T_{g}^{(\rho)}$, of the group $S(M)$ whose covariant transformations,
\begin{equation}\label{TgA}
(T_{g}^{(\rho)}\psi_{(\rho)})[\phi_{g}(x)]=\rho[A_{g}(x)]\psi_{(\rho)}(x)\,,
\end{equation}
leave the field equation invariant since their basis-generators
\cite{ES},
\begin{equation}\label{Xa}
X_{a}^{(\rho)}=i{\partial_{\xi^a} T_{g(\xi)}^{(\rho)}}_{|\xi=0}=-i
k_a^{\mu}\partial_{\mu} +\frac{1}{2}\,\Omega^{\hat\alpha\hat\beta}_{a}
\rho(S_{\hat\alpha\hat\beta})\,,
\end{equation}
commute with the operator of the field equation. Moreover, these generators  satisfy the commutation rules $[X_{a}^{(\rho)},
X_{b}^{(\rho)}]=ic_{abc}X_{c}^{(\rho)}$ determined by the structure constants, $c_{abc}$, of the algebras $s(M)\sim i(M)$. In other words, they are the basis-generators of a CR of the $s(M)$ algebra {\em induced} by the representation $\rho$ of the $sl(2,{\Bbb C})$ algebra. These generators can be put in (general relativistic) covariant form either in non-holonomic frames \cite{ES} or even in holonomic ones \cite{EPL}, generalizing thus the formula given by Carter and McLenaghan for the Dirac field \cite{CML}.

The generators (\ref{Xa}) have, in general, point-dependent spin terms which do not commute with the orbital parts. However, there are tetrad-gauges in which at least the generators of a subgroup  $H \subset I(M)$ may have
point-independent spin terms commuting with the orbital parts. Then
we say that the restriction to $H$ of the CR $T^{(\rho)}$ is {\em
manifest} covariant \cite{ES}. Obviously, if $H=I(M)$ then the whole
representation $T^{(\rho)}$ is manifest covariant. In particular, the linear CRs on the Minkowski spacetime have this property.

\section{The de Sitter isometries}

Let $(M,g)$ be the de Sitter spacetime defined as the
hyperboloid of radius $1/\omega$ \footnote{We denote by $\omega$
the Hubble de Sitter constant since  $H$ is reserved for the energy operator} in the five-dimensional flat spacetime $(M^5,\eta^5)$ of coordinates $z^A$  (labeled by the indices $A,\,B,...= 0,1,2,3,4$) and metric $\eta^5={\rm diag}(1,-1,-1,-1,-1)$. The local charts $\{x\}$  can be introduced on $(M,g)$ giving the set of functions $z^A(x)$ which solve the hyperboloid equation,
\begin{equation}\label{hip}
\eta^5_{AB}z^A(x) z^B(x)=-\frac{1}{\omega^2}\,.
\end{equation}
Here we use the chart $\{t,\vec{x}\}$ with the conformal time $t$ and Cartesian spaces coordinates $x^i$ defined by
\begin{eqnarray}
z^0(x)&=&-\frac{1}{2\omega^2 t}\left[1-\omega^2({t}^2 - \vec{x}^2)\right]
\nonumber\\
z^i(x)&=&-\frac{1}{\omega t}x^i \,, \label{Zx}\\
z^4(x)&=&-\frac{1}{2\omega^2 t}\left[1+\omega^2({t}^2 - \vec{x}^2)\right]
\nonumber
\end{eqnarray}
This chart  covers the expanding part of $M$ for $t \in (-\infty,0)$
and $\vec{x}\in {\Bbb R}^3$ while the collapsing part is covered by
a similar chart with $t >0$. Both these charts have the
conformal flat line element,
\begin{equation}\label{mconf}
ds^{2}=\eta^5_{AB}dz^A(x)dz^B(x)=\frac{1}{\omega^2 {t}^2}\left({dt}^{2}-d\vec{x}^2\right)\,.
\end{equation}
In addition, we consider the local frames $\{t,\vec{x};e\}$ of the diagonal gauge,
\begin{equation}\label{tt}
e^{0}_{0}=-\omega t\,, \quad e^{i}_{j}=-\delta^{i}_{j}\,\omega t
\,,\quad
\hat e^{0}_{0}=-\frac{1}{\omega t}\,, \quad \hat e^{i}_{j}=-\delta^{i}_{j}\,
\frac{1}{\omega t}\,.
\end{equation} 

The gauge group $G(\eta^5)=SO(1,4)$  is the isometry group of $M$,  since its transformations, $z\to gz$,  $g\in SO(1,4)$, leave the equation (\ref{hip}) invariant. Its universal covering group ${\rm Spin}(\eta^5)=Sp(2,2)$ is not involved directly in our construction since the spinor CRs are induced by the spinor representations of its subgroup $SL(2,\Bbb C)$. Therefore, we can restrict ourselves to the group $SO(1,4)$ for which we adopt the parametrization
\begin{equation}
g(\xi)=\exp\left(-\frac{i}{2}\,\xi^{AB}\sigma_{AB}\right)\in SO(1,4) 
\end{equation}
with skew-symmetric parameters, $\xi^{AB}=-\xi^{BA}$,  and the covariant generators $\sigma_{AB}$ of the fundamental representation of the $so(1,4)$ algebra carried by $M^5$. These have the matrix elements, 
\begin{equation}
(\sigma_{AB})^{C\,\cdot}_{\cdot\,D}=i\left(\delta^C_A\, \eta_{BD}
-\delta^C_B\, \eta_{AD}\right)\,.
\end{equation}
The principal $so(1,4)$ generators with physical meaning \cite{CCC} are the energy $\hat h=\omega\sigma_{04}$, angular momentum  $\hat j_k=\frac{1}{2}\varepsilon_{kij}\sigma_{ij}$, Lorentz boosts $\hat k_i=\sigma_{0i}$, and the Runge-Lenz-type vector $\hat r_i=\sigma_{i4}$. In addition, it is convenient to introduce the momentum $\hat p_i=-\omega(\hat r_i+\hat k_i)$ and its dual $\hat q_i=\omega(\hat r_i-\hat k_i)$ which are nilpotent matrices ($\hat p_i^3=\hat q_i^3=0$) generating two Abelian three-dimensional subalgebras, $t(3)_P$ and respectively $t(3)_Q$. All these generators may form different bases of the algebra $so(1,4)$ as, for example, the basis $\{\hat h,\hat p_i,\hat q_i,\hat j_i\}$ or the Poincar\' e-type one, $\{\hat h,\hat p_i,\hat j_i,\hat k_i\}$. We note that the four-dimensional restriction, $\{j_i,k_i\}$, of the $so(1,3)$ subalegra generate the vector representation of the group $L_+^{\uparrow}$.

Using these generators we can derive the $SO(1,4)$ isometries, $\phi_{g}$, defined as  
\begin{equation}\label{zgz}
z[\phi_{g}(x)]=g\,z(x). 
\end{equation}
The transformations $g\in SO(3)\subset SO(4,1)$  generated by $\hat j_i$, are simple rotations of $z^i$ and $x^i$ which transform alike since this symmetry is global. The transformations generated by $\hat h$,
\begin{equation}\label{transH}
\exp(-i\xi \hat h)\,:\quad
\begin{array}{lcl}
z^0&\to&z^0 \cosh\alpha-z^4 \sinh\alpha \\
z^i&\to&z^i\\
z^4&\to&-z^0 \sinh\alpha+z^4 \cosh\alpha 
\end{array}
\end{equation}
whith $\alpha=\omega\xi$, produce the dilatations
$t\to t\,e^{\alpha}$ and $x^i\to x^i e^{\alpha}$,
while the $t(3)_P$ transformations
\begin{equation}\label{transP}
\exp(-i{\xi}^i\hat p_i)\,:\quad
\begin{array}{lcl}
z^0&\to&z^0 +\omega\,\vec{\xi}\cdot\vec{z}+\frac{1}{2}\,\omega^2
{\vec{\xi}\,}^2\,(z^0+z^4) \\
z^i&\to&z^i+\omega\,\xi^i\,(z^0+z^4)\\
z^4&\to&z^4 -\omega\,\vec{\xi}\cdot\vec{z}-\frac{1}{2}\,\omega^2
{\vec{\xi}\,}^2\,(z^0+z^4) 
\end{array}
\end{equation}
give rise to the space translations $ x^i\to x^i +\xi^i$ at fixed $t$.
More interesting are the $t(3)_Q$ transformations generated by $\hat q_i/\omega$,
\begin{equation}\label{transN}
\exp(-i{\xi}^i \hat q_i/\omega)\,:\quad
\begin{array}{lcl}
z^0&\to&z^0 -\vec{\xi}\cdot\vec{z}+\frac{1}{2}\,
{\vec{\xi}\,}^2\,(z^0-z^4) \\
z^i&\to&z^i-\xi^i\,(z^0-z^4)\\
z^4&\to&z^4 -\vec{\xi}\cdot\vec{z}+\frac{1}{2}\,
{\vec{\xi}\,}^2\,(z^0-z^4) 
\end{array}
\end{equation}
which lead to the isometries
\begin{eqnarray}
t&\to&\frac{t}{1-2\omega\, \vec{\xi}\cdot\vec{x}
-\omega^2{\vec{\xi}\,}^2\,({t}^2-\vec{x}^2)} \\
x^i&\to&\frac{x^i+\omega\xi^i\, ({t}^2-\vec{x}^2)}
{1-2\omega\, \vec{\xi}\cdot\vec{x}
-\omega^2{\vec{\xi}\,}^2\,({t}^2-\vec{x}^2)}\,.
\end{eqnarray}
We observe that $z^0+z^4=-\frac{1}{\omega^2 t}$ is invariant under translations (\ref{transP}), fixing the value of $t$, while $z^0-z^4=\frac{t^2-\vec{x}^2}{t}$ is left unchanged by the $t(3)_Q$ transformations (\ref{transN}). 

\section{Nachtmann's covariant representations}

In this context Nachtmann proposed another method of deriving covariant representations induced by the group $L^{\uparrow}_{+}$ that works in the case of  hyperbolic manifolds. We present this method of constructing NCRs using our formalism and denoting for brevity $G\equiv G(\eta)=L^{\uparrow}_{+}$ and $G_5\equiv G(\eta^5)=SO(1,4)$. The idea is to apply the Wigner theory of  induced representations but in configurations instead of the momentum representation \cite{Nach}.

This can be done since the de Sitter manifold is isomorphic to the space of left cosets $G_5/G$. Indeed, if one fixes the point $z_o=(0,0,0,0,\omega^{-1})^T\in M$, of local coordinates $(-\omega^{-1},0,0,0)$, then the whole de Sitter manifold is the orbit $M=\{g z_o | g\in G_5/G\} \subset M^5$ since the subgroup $G$ is just the stable group of $z_o$ ($gz_o=z_o\,,\, \forall g\in G$). Then any point $z(x)\in M$ can be reached  performing the 'boost' transformation $b(x):z_o\to z(x)=b(x)z_o$ whose matrix \cite{Nach},
\begin{equation}
b(x)= \exp(-ix^i\hat p_i) \exp(-i\alpha \hat h)\,, \quad \alpha=\ln(-\omega t)\,,   
\end{equation}
has the form 
\begin{equation}\label{boost}
b(x)=\left(
\begin{array}{ccccc}
-\frac{1+\omega^2(t^2+\vec{x}^2)}{2\omega t}&\omega x^1&\omega x^2&\omega x^3&-\frac{1-\omega^2(t^2-\vec{x}^2)}{2\omega t}\\
-\frac{x^1}{t}&1&0&0&-\frac{x^1}{t}\\
-\frac{x^2}{t}&0&1&0&-\frac{x^2}{t}\\
-\frac{x^3}{t}&0&0&1&-\frac{x^3}{t}\\
-\frac{1-\omega^2(t^2+\vec{x}^2)}{2\omega t}&-\omega x^1&-\omega x^2&-\omega x^3&-\frac{1+\omega^2(t^2-\vec{x}^2)}{2\omega t}\\
\end{array}\right)
\end{equation}
as it results from equations (\ref{transH}) and (\ref{transP}). Furthermore, one can verify that the {\em canonical} five-dimensional 1-forms $\tilde\omega_5(x)=b^{-1}(x)d\,b(x)\,z_o$ which satisfy $ds^2=\eta^5_{AB}\tilde\omega_5^A(x)\tilde\omega_5^B(x)$ define just the tetrad-gauge (\ref{tt}) since,
\begin{equation}
\tilde\omega_5^{\hat\alpha}(x)=\hat e^{\hat\alpha}_{\mu}(x)dx^{\mu}\,,\quad 
\tilde\omega_5^{4}(x)=0\,.
\end{equation}  
The boosts (\ref{boost}) are defined up to an arbitrary gauge, $b(x)\to b(x)\lambda^{-1}(x)$, $\lambda(x)\in G$, that transform the  1-forms  as $\tilde\omega_5(x) \to \lambda(x)\,\tilde\omega_5(x)$. It remains to define the NCRs according to which  the covariant fields defined on $M$ transform under isometries.  

Let us start with the field $\psi^5 :M^5\to {\cal V}$ which transforms according to the representation $g\to T_{g}^5$ whose action, $(T^5_g \psi^5)(gz)=\rho_5(g)\psi^5(z)$, is given by the finite-dimensional representation $\rho_5$ of $G_5$, carried by the vector space ${\cal V}$. Then one can define the field $\psi:M \to {\cal V}$ assuming that \cite{Nach}
\begin{equation}
\psi(x)=\left(T^5_{b^{-1}(x)}\psi^5 \right)(z_o)=\rho_5[b^{-1}(x)]\psi^5[z(x)]\,.
\end{equation} 
The last step is to define the NCRs $g\to T_g$ as 
\begin{equation}
(T_g \psi)(x)=\rho_5[b^{-1}(x)]\left(T^5_g\psi^5\right)[z(x)]\,,
\end{equation}
obtaining straightforwardly the transformation rule under isometries,
\begin{equation}\label{NCR}
(T_{g}\psi)[\phi_{g}(x)]=\rho[\lambda_{g}(x)]\psi(x)\,, 
\end{equation}
where $\phi_{g}$ is defined by equation (\ref{zgz})  while $\rho=\rho_5|_G$ is the restriction to $G$ of the representation $\rho_5$ since \cite{Nach}
\begin{equation}\label{WNCR}
\lambda_{g}(x)=b^{-1}[\phi_{g}(x)]g\,b(x)\in G \,.
\end{equation}
Hereby it results that the NCRs are {\em induced} by the finite-dimensional representations of the same group, $G=L^{\uparrow}_{+}$, just as our CRs. 

Now the question we may ask is how these two types of representations are are related to each other. The solution is very simple if we observe that the identity $d b[\phi_{g}(x)]z_o=g\,d b(x)z_o$, derived from equation (\ref{zgz}), enables us to write the transformation rule 
$\tilde\omega_5[\phi_{g}(x)]=\lambda_{g}(x)\tilde\omega_5(x)$ which shows that the NCRs are {\em  equivalent} to CRs in the local frames $\{t,\vec{x};e\}$ of the de Sitter manifold.  Moreover, the four dimensional restriction of the matrix
\begin{equation}
\lambda_{g}(x)=\left(
\begin{array}{cc}
\Lambda[A_{g}(x)]&0\\
0&1
\end{array}\right)
\end{equation}
is just the transformation matrix  (\ref{Axx}) in the frame $\{t,\vec{x};e\}$.

This equivalence testifies the consistence of the theory of covariant fields in the local frames of curved manifolds. Moreover, the Nachtmann method on the de Sitter spacetime represents a useful tool that completes our general theory of  CRs. In the next section we give examples of induced transformations calculated using both these methods. 

\section{Transformation rules}

We have seen that the NCRs$\sim$CRs are induced representations that differ from the usual linear ones apart from the scalar case when all of them reduce to the natural representation.  Therefore, it deserves to present some details concerning the transformation rules of the CRs which are not discussed so far.

\subsection{Continuous transformations}

The components of the covariant fields, $\psi_{(\rho)}:\, M\to {\cal V}_{(\rho)}$, in local frames $\{t,\vec{x};e\}$, transform according to the CRs (\ref{TgA}) which can be written now as   
\begin{equation}\label{TgP}
T_{g}^{(\rho)}\psi_{(\rho)}=[\rho(A_{g})\psi_{(\rho)}]\circ \phi_g^{-1}\,,
\end{equation}
while their basis-generators defined by equation (\ref{Xa}) take the form,
\begin{equation}\label{XadS}
X_{(AB)}^{(\rho)}=-i
k_{(AB)}^{\mu}\partial_{\mu} +\frac{1}{2}\,\Omega^{\hat\alpha\hat\beta}_{(AB)}
S_{\hat\alpha\hat\beta}^{(\rho)}\,,
\end{equation}
where we denote $S_{\hat\mu\hat\nu}^{(\rho)}$  instead of  $\rho(S_{\hat\mu\hat\nu})$ bearing in mind that the functions (\ref{Om}) have to be calculated using the Killing vectors $k_{(AB)}$ corresponding to the parameters $\xi^{AB}$. These generators form an operator-valued CR of the $so(1,4)$ algebra induced by $\rho$. The basis under consideration here is constituted by the energy $H$, momentum $P_i$, angular momentum $J_i^{(\rho)}$ and $Q_i^{(\rho)}$ operators that read \cite{CCC},
\begin{eqnarray}
H &\equiv &  \omega X_{(04)}=-i\omega(t\,\partial_t+ {x}^i {\partial}_i)\,,\\
P_i&\equiv & \omega(X_{(i0)}-X_{(i4)})=-i\partial_i\,,\label{Pi}\\
J_i^{(\rho)}&\equiv & \textstyle{\frac{1}{2}}\varepsilon_{ijk}X_{(jk)}=\varepsilon_{ijk}(-ix^j\partial_k+\frac{1}{2}S_{jk}^{(\rho)})\,,\\
Q_i^{(\rho)} &\equiv & \omega(X_{(i0)}+X_{(i4)})=-2i \omega^2 x^i H+
\omega^2({\vec{x}\,}^2-t^2)P_i\nonumber\\
&&\hspace*{34mm}-2\omega^2(S^{(\rho)}_{i0}t+S^{(\rho)}_{ij}x^j)\,,\label{Qi}
\end{eqnarray}
We studied the properties of this type of CRs deriving their Casimir operators whose eigenvalues indicate that any subspace ${\cal V}_s \subset {\cal V}_{\rho}$ of given spin $s$  corresponds to a unitary irreducible representation of the $so(1,4)$ algebra which is determined by the rest energy and spin \cite{CCC}.

All this machinery is based on the form of the induced   transformations $\Lambda_g$ which was never discussed in literature. For this reason we briefly review some of their properties in the local frame $\{t,\vec{x};e\}$ where we consider the isometries generated by $\{\hat h,\hat p_i,\hat q_i,\hat j_i\}$. From equation (\ref{TgP}) we see that the interesting piece is $\hat\Lambda_g=\Lambda_g\circ \phi_g^{-1}$ that has to be calculated as the four-dimensional restriction of the matrix 
\begin{equation}
\hat\lambda_g= \lambda_g\circ\phi^{-1}_g=b^{-1}g (b\circ\phi^{-1}_g)
\end{equation}
resulted from equation (\ref{WNCR}). First of all we find that the transformations $g$ generated by $\hat h$ or $\hat p_i$ give $\hat \Lambda_g=I$. This explains why the corresponding generators, $H$ and $P_i$,  are the only generators without spin terms.  The rotations generated by $\hat j_i$ form a global symmetry so that 
\begin{equation}
\hat \Lambda_g=\left(
\begin{array}{cc}
1&0\\
0&R(g)
\end{array}\right)\,,\quad \forall \,g \in SO(3)\subset SO(1,4)\,, 
\end{equation}
where $R(g)\in SO(3)\subset L^{\uparrow}_+$ denotes the usual point-independent rotation matrices. This is related to the fact that the rotation generators $J_i^{(\rho)}$ have usual spin terms, independent on coordinates \cite{CCC}.

The other transformations generated by $\hat q_i$,  $\hat k_i$ and $\hat r_i$  induce complicated point-dependent transformation matrices $\hat\Lambda_g(x)$. For example, the transformation $g_1\in t(3)_Q$ parametrized  as $g_1=\exp(-i  \hat q_1 \xi/\omega^2)$ gives rise to the matrix
\begin{equation}
\hat\Lambda_{g_1}(x)=\frac{1}{f(\xi,x)}\left\{ I+2\xi \alpha_1(x) + 2\xi^2 [\alpha_1(x)^2+x^1\alpha_1(x)]
\right\}\,, 
\end{equation}
where
\begin{eqnarray}
\alpha_1(x)=i(k_1 t+j_2 x^3-j_3 x^2)=\left(
\begin{array}{cccc}
0&t&0&0\\
t&0&-x^2&-x^3\\
0&x^2&0&0\\
0&x^3&0&0
\end{array}\right)
\end{eqnarray} 
and
\begin{equation}
f(\xi,x)=1+2\xi x^1-\xi^2(t^2-\vec{x}^2)\,.
\end{equation}
As mentioned above, the matrices $j_i$ and $k_i$ are the generators of the vector representation of the group $L^{\uparrow}_+$.

\subsection{Discrete transformations}

The discrete symmetries on $M$ were less studied  since the mirror transformations defined directly on $M$ depend on the concrete choice of the local coordinates and this fact could lead to some ambiguities. Another possibility is to consider the discrete transformations on $M^5$ inducing isometries on $M$. This method was  used for linear representations \cite{disc} but never for CRs. For this reason we focus on the transformations  produced by the discrete symmetries.

The simplest $O(1,4)$ discrete transformations on $M^5$, denoted by
${\pi_{[A]}}$, are those which change the sign of a single
coordinate, $z^A\to -z^A$. These transformations give rise to the
discrete isometries $x\to x'=\phi_{[A]}(x)$ that, according to equation(\ref{zgz}), are defined as $z[\phi_{[A]}(x)]=\pi_{[A]}z(x)$. Obviously, these isometries satisfy $\phi_{[A]}\circ\phi_{[A]}=id$. A rapid inspection indicates that the interesting non-trivial isometries are produced by $\pi_{[0]}$,
\begin{equation}\label{Pi0}
t'=\phi_{[0]}^0(x)=\frac{t}{\omega^2 (t^2-{\vec{x}\,}^2)}\,, \quad
x^{i\,\prime}=\phi_{[0]}^i(x)=\frac{x^i}{\omega^2
(t^2-{\vec{x}\,}^2)}\,,
\end{equation}
and by $\pi_{[4]}$ which gives
\begin{equation}\label{Pi4}
t'=\phi_{[4]}^0(x)=-\frac{t}{\omega^2 (t^2-{\vec{x}\,}^2)}\,, \quad
x^{i\,\prime}=\phi_{[4]}^i(x)=-\frac{x^i}{\omega^2
(t^2-{\vec{x}\,}^2)}\,.
\end{equation}
The other isometries, $\phi_{[i]}$, are simple mirror transformations
of the space coordinates $x^i$ so that the (space) parity reads
$\phi_{(\vec{x})}=\phi_{[1]}\circ\phi_{[2]}\circ\phi_{[3]}$. Another
remarkable discrete isometry is $\phi_{(x)}=\phi_{[0]}\circ\phi_{[4]}$
which changes the signs of all the coordinates $x^{\mu}$. The antipodal transformation $z\to -z$ gives rise to the isometry $\phi_{(t)}=\phi_{(x)}\circ\phi_{(\vec{x})}$ that plays the role of the time reversal on $M$ changing $t\to -t$.

It is worth pointing out that the physical measurements can be performed only
inside the light-cone where $|t|>|\vec{x}|$. This means that the
isometry (\ref{Pi0}) does not change the signs of the time and space
coordinates and, consequently, the charts $\{t,\vec{x}\}$ and
$\{t',{\vec{x}\,}'\}$ cover the same portion of $M$. On the
contrary, the isometry (\ref{Pi4}) changes the sign of the conformal
time moving the transformed chart to the opposite portion of this
manifold. The antipodal isometry also changes these portions between themselves.

The last step is to find the transformation rules of the covariant fields calculating the four-dimensional restriction of the matrix (\ref{WNCR}). The surprise is that for both the isometries, $\phi_{[0]}$ and $\phi_{[4]}$, we obtain the same result,
\begin{equation}
\Lambda_{[0]}(x)=\Lambda_{[4]}(x)=\exp\left(i k_i\,\frac{x^i}{|\vec{x}|}\, \alpha\right)\,T \,,\quad \alpha={\rm arctanh}\left(\frac{2 t |\vec{x}|}{t^2+\vec{x}^2}\right)\,,
\end{equation}
where $T={\rm diag}(-1,1,1,1)\in O(1,3)$ is the usual time reversal matrix. For the parity transformation we find $\Lambda_{(\vec{x})}=P={\rm diag}(1,-1,-1,-1)\in O(1,3)$. The other discrete transformations can be obtained combining these presented here. 

\section{Concluding remarks}

The fact that the  Nachtmann approach and our general theory of external symmetry lead to the same results in the de Sitter case, convince us that we have a coherent theory of induced CRs according to which the covariant fields with spin must transform under isometries. The generators of these representations are the conserved quantum observables that enable one to derive quantum modes as common eigenstates of suitable systems of commuting operators. Thus the theory of CRs presented here could be a step forward to a theory of covariant fields canonically quantized on the de Sitter spacetime.

Mathematically speaking, the problem which remains open is the equivalence among the CRs and the unitary irreducible representations of the group $Sp(2,2)$ \cite{UIR}. As mentioned above, our study of the Casimir operators of the de Sitter isometry group \cite{CCC} indicates that any subspace of fixed spin of the spaces carrying  CRs  corresponds to a unitary irreducible representation of the algebra $sp(2,2)\sim so(1,4)$. However, we do not have yet the concrete transformations able to realize this equivalence at the level of the group representations as it happens in the flat case with the usual CRs on the Minkowski spacetime and the unitary representations of the Poincar\' e group \cite{WKT}. We hope to make some progress in solving this problem by combining both the methods of constructing CRs discussed above.

\end{document}